\documentclass{article}

\usepackage{PRIMEarxiv}

\usepackage{comment}
\usepackage[utf8]{inputenc}         
\usepackage[T1]{fontenc}            
\usepackage[hidelinks]{hyperref}    
\usepackage{url}                    
\usepackage{booktabs}               
\usepackage{amsfonts}               
\usepackage{nicefrac}               
\usepackage{microtype}              
\usepackage{lipsum}
\usepackage{fancyhdr}               
\usepackage{graphicx}               
\graphicspath{{media/}}             

\title{PIM-AI: A Novel Architecture for High-Efficiency LLM Inference
}

\author{
  Cristobal Ortega, Yann Falevoz, Renaud Ayrignac \\
  UPMEM \\
  Grenoble \\
  \texttt{contact@upmem.com} \\
}

\begin{document}
\maketitle

\begin{abstract}
Large Language Models (LLMs) have become essential in a variety of applications due to their advanced language understanding and generation capabilities. However, their computational and memory requirements pose significant challenges to traditional hardware architectures. Processing-in-Memory (PIM), which integrates computational units directly into memory chips, offers several advantages for LLM inference, including reduced data transfer bottlenecks and improved power efficiency.

This paper introduces PIM-AI, a novel DDR5/LPDDR5 PIM architecture designed for LLM inference without modifying the memory controller or DDR/LPDDR memory PHY. We have developed a simulator to evaluate the performance of PIM-AI in various scenarios and demonstrate its significant advantages over conventional architectures. In cloud-based scenarios, PIM-AI reduces the 3-year TCO per queries-per-second by up to 6.94x compared to state-of-the-art GPUs, depending on the LLM model used. In mobile scenarios, PIM-AI achieves a 10- to 20-fold reduction in energy per token compared to state-of-the-art mobile SoCs, resulting in 25 to 45~\% more queries per second and 6.9x to 13.4x less energy per query, extending battery life and enabling more inferences per charge. 

These results highlight PIM-AI's potential to revolutionize LLM deployments, making them more efficient, scalable, and sustainable.
\end{abstract}


\section{Introduction}
\subsection{LLM Applications and Rapid Evolution}
Large Language Models (LLMs) have become an integral part of many domains~\cite{A_Comprehensive_Overview_of_Large_Language_Models, A_Survey_of_Large_Language_Models}, including healthcare, education, social media, business, law, creative industries, and scientific research. They demonstrate exceptional capabilities in natural language processing (NLP) tasks such as language translation, text generation, and question answering~\cite{A_Review_on_Large_Language_Models}. Their rapid development has been driven by advances in deep learning, increased computational resources, and large training datasets. LLMs now play a critical role in achieving human-like literacy and communication in machines, addressing a long-standing challenge in artificial intelligence (AI).

LLMs evolved from statistical approaches and n-gram models~\cite{n-gram}, which struggled with long-term dependencies~\cite{A_Review_on_Large_Language_Models}. RNNs improved sequential data modeling~\cite{RNN_based_language_model} but had issues like vanishing gradients~\cite{A_Review_on_Large_Language_Models}. The breakthrough came with the Transformer architecture~\cite{Attention_is_all_you_need}, which used self-attention to handle long-range dependencies effectively. This led to models like BERT~\cite{BERT} and GPT~\cite{GPT1}. Encoder-decoder models like T5~\cite{T5} and BART~\cite{BART} unified understanding and generation tasks. LLMs also developed emergent capabilities, such as in-context learning seen in GPT-3~\cite{GPT3}. Specialized models like Codex~\cite{Codex} and WebGPT~\cite{WebGPT} showed diverse applications. Recent models, including Meta's LLaMA~\cite{LLaMA, Llama2}, Google’s PaLM~\cite{PaLM, PaLM_2} and Mistral 7B~\cite{Mistral_7B}, exhibit strong performance with fewer parameters. Guided by scaling laws, LLMs like GPT-4~\cite{GPT4} and Mixtral-8x22B~\cite{Mistral_8x22B} achieve high coherence and relevance across tasks, pushing the boundaries of AI.

The continued evolution and versatility of LLMs underscores their transformative impact as they continue to integrate into both professional and everyday contexts.

\subsection{Key Elements and Architectures of LLMs}
Large Language Models (LLMs) perform inference by transforming input words or prompts into tokens, which are the basic units of these models. Tokens can represent words or characters~\cite{Tokenization}. The tokenized input is then processed by the model.

\begin{figure}[!htb]
\centering
\includegraphics[width=0.65\columnwidth]{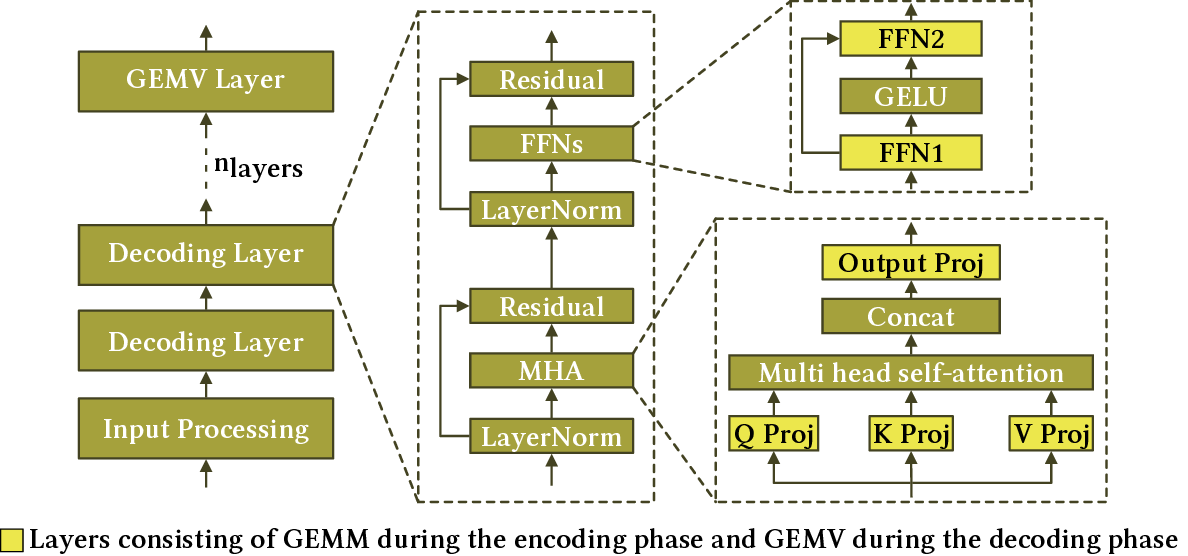}
\caption{Simplified architecture of transformer-based LLMs, showing the common structure used in both encoding and decoding phases.} \label{Architecture}
\end{figure}

The execution of LLM is divided into two phases: encoding and decoding. Figure~\ref{Architecture} shows the structure of these phases, which use the same basic structure. Tokens are first translated into embeddings, which are numerical representations of the tokens~\cite{Distributed_Representations_Words_Phrases, Glove}. These embeddings pass through layers consisting of an input normalization layer~\cite{Normalization, Layer_Normalization}, a~multi-head attention (MHA) mechanism, an output linear projection, and a feed-forward layer~\cite{Attention_is_all_you_need}. After the last layer, the embeddings are translated back into tokens and the next token is selected.

The MHA mechanism computes the attention of each token with respect to all previous tokens and includes three components for each token: \textit{Query} (Q), \textit{Key} (K), and \textit{Value} (V). To reduce redundant computations, the KV cache stores previously computed keys and values, offloading some of the computational load to memory~\cite{KVCache}.

The balance between memory bandwidth and compute performance depends on the size of the input matrix (embeddings)~\cite{Dissecting_batching_effects}. The balance point varies from a few tens to a few hundred tokens, depending on the hardware.

During the encoding phase, the entire initial prompt is processed. The yellow layers in Figure~\ref{Architecture} are general matrix multiplications (GEMM), where the input matrix has as many rows as there are tokens in the prompt. This makes the encoding compute-bound, since the full prompt is typically large enough to require significant computational resources. The KV cache is populated based on the parameters of the encoders/decoders and the MHA. Once the initial prompt is processed, the model generates a token that marks the transition to the decoding phase.

In the decoding phase, the most recently generated token is processed using the KV cache. Since the yellow layers in Figure~\ref{Architecture} process a single row matrix, i.e. general matrix vector multiplications (GEMV), the decoding is memory-bound. New keys and values are appended to the KV cache. This cycle of token generation and decoding continues until a stop token is generated or the maximum number of tokens is reached.

To optimize decoding performance, it is beneficial to batch multiple requests to balance memory bandwidth and compute performance~\cite{Dissecting_batching_effects}. This technique is common in cloud environments, but is not always feasible for mobile or edge devices, which typically run on more constrained systems.

\subsection{Challenges of Executing LLMs}
The vast majority of LLMs today run in the cloud~\cite{MobileLLM, Octopus}, which presents significant challenges due to their compute- and memory-intensive nature. Scalability and cost issues arise from the need to support large numbers of users, requiring significant compute resources~\cite{A_Survey_on_Hardware_Accelerators_for_LLMs} and resulting in high operational costs~\cite{Energy_and_Policy_Considerations_for_Deep_Learning_in_NLP}. Real-time applications, such as virtual assistants, suffer from high latency due to data transfer times and processing delays. The memory wall problem~\cite{Hitting_the_memory_wall, AI_and_Memory_Wall}, caused by the separation of memory and processing units, leads to inefficiencies due to slow data transfers, especially for large \textit{Key-Value} caches. Running LLMs in the cloud using GPUs consumes significant energy, increasing both costs and environmental concerns~\cite{From_Words_to_Watts, Measuring_and_Improving_the_Energy_Efficiency_of_Large_Language_Models_Inference, Energy_and_Policy_Considerations_for_Deep_Learning_in_NLP, Risks_and_benefits}. In addition, processing sensitive data in the cloud poses security and privacy risks~\cite{Viewing_LLMs_in_Legal_Aspect, LLM_to_preserve_privacy, Preserving_Privacy}, making it difficult to comply with data protection regulations. Maintaining the necessary infrastructure is complex and resource-intensive, requiring regular updates and skilled personnel~\cite{Deliver_high_performance_ML}.

Given these challenges, there is growing interest in deploying LLMs on mobile and edge devices to provide more sustainable, cost-effective, and accessible AI solutions that can operate offline or with low connectivity~\cite{MobileLLM, MM1, Octopus}. This approach also addresses security and privacy concerns by keeping data processing local to the device. However, there are several barriers to overcome. LLMs such as LLaMA-13B, which require approximately 26 GB of memory in 16-bit data types, cannot be efficiently deployed on current mobile SoCs (systems on chip). Mobile devices have limited battery life, making it difficult to deploy energy-intensive LLMs without significant optimizations. In addition, prolonged operation of high-performance models can lead to overheating, impacting device performance and longevity. 

Overcoming these challenges is critical to the efficient and effective use of LLMs in various applications, both in the cloud and on mobile platforms.

\subsection{Rationale for a Novel PIM Architecture}
Researchers and engineers are exploring a variety of solutions to improve the efficiency and scalability of LLMs and overcome the computational and storage challenges associated with running LLMs.

\subsubsection{Model Reduction Techniques:}
Several techniques have been developed to make LLMs less compute- and memory-intensive. Quantization reduces model weights to smaller bit widths to save memory and energy~\cite{SmoothQuant, Compression_of_Generative_Pre-trained_LLMs_via_Quantization, Survey_of_Quantization_Methods, OneBit, The_Era_of_1-bit_LLMs} but can lead to a loss of accuracy at very low bit widths. Pruning~\cite{LLM-Pruner, From_Dense_to_Sparse} removes unimportant weights to enhance efficiency, though it may degrade performance by eliminating critical parts of the model. Compression~\cite{Survey_on_Model_Compression_for_LLMs} reduces model size but can slow down inference due to the need to decompress data on the fly. Knowledge distillation~\cite{Distilling_Knowledge_in_NN, Knowledge_distillation_survey} trains smaller models under the guidance of larger ones but often requires extensive fine-tuning and may not match the performance of their larger counterparts. 

\subsubsection{Traditional Hardware Accelerators:}
An alternative approach to model reduction is the development of specialized hardware accelerators~\cite{A_Survey_on_Hardware_Accelerators_for_LLMs}. Field Programmable Gate Arrays (FPGAs) offer flexibility through reprogrammability, allowing customization for specific workloads. However, they are often expensive and require significant expertise to program efficiently. CPUs and GPUs excel at parallel processing and are widely used for AI tasks, but they face significant inefficiencies. GPUs in particular have high power consumption and cost. Application-Specific Integrated Circuits (ASICs) offer the highest performance and energy efficiency for fixed tasks, but are extremely expensive to develop and lack flexibility for different models.

Despite their advantages, these traditional accelerators are based on the von Neumann architecture where processing and memory are separated.
This leads to inefficiencies due to the memory wall problem~\cite{Hitting_the_memory_wall, AI_and_Memory_Wall}, where the speed at which data can be transferred between memory and processing units lags behind processor speeds. This mismatch causes significant delays and energy consumption, especially during attention computations in LLMs.

While high-speed memory technologies such as High Bandwidth Memory (HBM) and Compute Express Link (CXL) improve capacity and bandwidth~\cite{Landscape_of_CNM_and_CIM}, they do not fully address the fundamental problem of data transfer bottlenecks.

\subsubsection{Processing-in-Memory (PIM) Accelerators:}
The PIM architecture integrates computational units into the memory chip itself, enabling data processing directly where the data is stored~\cite{ModernPrimerOnPIM}. This approach eliminates the need for extensive data transfers between memory and CPU/GPU, directly addressing the memory wall issue. By exploiting bank parallelism within the DRAM structure, PIM can achieve internal bandwidths many times greater than external bandwidths.

Emerging memory technologies~\cite{High-speed_emerging_memories_for_AI_hardware_accelerators}, such as capacitorless gain cell-based eDRAM, ferroelectric memory, spin-transfer torque magnetic random access memory (STT-MRAM), and spin-orbit torque magnetic random access memory (SOT-MRAM), are being explored but are far from commercialization. More mature technologies, such as UPMEM PIM~\cite{UPMEMTechPaper} and Samsung's HBM PIM~\cite{Samsung_HBM_PIM}, seem more realistic as a short-term solution:

\textbf{UPMEM PIM} is the first commercially available PIM architecture. It combines each DRAM bank with a custom 32-bit processor capable of running 24 threads independently. While UPMEM PIM shows impressive speedups, as well as cost and energy reduction for data-intensive workloads~\cite{PrIM, Energy_Efficiency_Impact_of_PIM}, its limitations in hardware-supported operations and overall processor performance make it unsuitable for handling the heavy computational load of LLMs.

\textbf{Samsung's HBM PIM} integrates floating-point SIMD (Single Instruction Multiple Data) units close to the memory banks, enabling high-speed, low-latency computations directly within the memory. Research~\cite{Samsung_HBM_PIM} has shown that this approach can significantly improve performance and energy efficiency on LLM inference tasks compared to traditional GPU-based systems.

PIM technology is a promising direction for improving the efficiency and scalability of LLMs, making it possible to deploy these powerful models in a more energy-efficient and cost-effective way. Ongoing research and development in this area continues to push the boundaries of what is possible, paving the way for more advanced and accessible AI applications.

\subsection{Research Objectives and Key Contributions}
This research introduces a novel accelerator architecture for LLMs and other memory-intensive workloads, overcoming limitations of traditional hardware accelerators. The core innovation is the PIM-AI architecture, integrating computational units directly into the memory chip, significantly reducing data transfer bottlenecks and improving overall performance and energy efficiency. 

A simulator was developed to analyze LLM execution on various hardware platforms, providing a comprehensive performance analysis of PIM-AI under different scenarios and highlighting its advantages over reference architectures. 

The paper is organized as follows: Section 2 introduces the PIM-AI architecture, Section 3 outlines the experimental setup and hardware simulator, Section 4 presents the results, Section 5 discusses the simulation results, limitations, and future research directions, and Section 6 concludes the paper.
\section{A new PIM architecture for LLMs}
The PIM-AI architecture is a novel PIM design aimed at addressing the computational and energy efficiency challenges of the encoding and decoding phases of LLM operations. This architecture integrates seamlessly with existing systems without requiring modifications to the memory controller or DDR/LPDDR memory PHY.

\subsection{PIM-AI Chip}
The PIM-AI chip operates in two modes: non-PIM mode and PIM mode. In non-PIM mode, the operating system~(OS) uses the full 2GB capacity of the memory chip as standard memory. In PIM mode, the chip acts as an accelerator, improving performance and energy efficiency.

\begin{figure}[!htb]
\centering
\includegraphics[width=0.7\columnwidth]{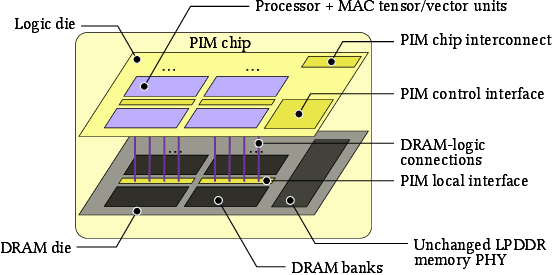}
\caption{PIM-AI chip architecture: The logic die houses 4 RISC V processors, each with tensor and vector units, accessing DRAM banks through DRAM-logic connections. The memory PHY remains unchanged.} \label{Chip}
\end{figure}

The PIM-AI architecture, shown in Figure~\ref{Chip}, features a stacked die configuration with a DRAM die and a logic die. The logic die contains four Linux-capable RISC-V processors, each equipped with tensor and vector units. These processors access DRAM banks at a bandwidth of up to 102.4~GB/s and a read/write energy consumption of 0.95 pJ/bit. The tensor units perform up to 8~TOPS (Tera Operations Per Second) and support data formats such as INT4, INT8, FP16, and BF16, using 32-bit registers for accumulation.

The Control Interface (CI) manages commands and responses between the host system (HOST) and the PIM-AI processor via specific physical addresses. Memory access is exclusive between the HOST and the PIM-AI processor, requiring the HOST to reclaim ownership for read/write operations. Due to hardware-level interleaving, each memory chip handles a subset of the data. A software driver manages the data transfers to ensure correct processing by each PIM-AI chip.

\subsection{PIM-AI DIMM} \label{ss:PIM-DIMM}
The PIM-AI architecture can be scaled by integrating multiple PIM-AI chips into a dual-in-line memory module~(DIMM), increasing both bandwidth and compute capacity. A typical PIM-AI DIMM contains 32GB of memory with a total aggregate bandwidth of 1.6TB/s and a compute capacity of up to 128~TFLOPs FP16/BF16.

Each PIM-AI DIMM contains an internal chip interconnect system that links all PIM-AI chips within the DIMM. This~interconnect facilitates data sharing among the chips, minimizing the data transfer load from the HOST and enabling effective parallel processing.

In a mode of operation where the same data needs to be sent to all chips, the HOST can partition the input data and send a segment of it to each PIM-AI chip. The chips then internally distribute their segments to all other PIM-AI chips within the DIMM, ensuring that each chip has access to the full input data set. This mechanism reduces the amount of data that must be written by the HOST.

In another mode of operation, where different data sets are processed by different PIM-AI chips, the internal communication system allows intermediate results to be exchanged between chips. This allows different stages of a processing pipeline to be parallelized, improving overall efficiency and performance.

During read operations, the HOST retrieves the results of the partial operations performed by each PIM-AI DIMM. This modular approach allows the PIM-AI architecture to efficiently handle large computations, taking advantage of the high bandwidth and processing power distributed across multiple DIMMs.
\section{Methodology}

\subsection{Hardware LLM Simulator}
To compare our proposed PIM-AI architecture with state-of-the-art hardware, we developed a simulator for PyTorch models. This simulator is available at \url{https://github.com/upmem/upmem_llm_framework}. It can execute multiple layers and functions during model inference without requiring modifications to the original PyTorch model. During~execution, the simulator collects metrics such as total number of TOPs (Tera Operations), execution time, data transfer sizes (H2D: HOST to device, D2H: device to HOST and main memory), energy consumption, and power consumption for a given hardware profile.

A hardware profile is a set of configurable parameters that represent a hardware accelerator. These parameters include: TOPS and energy per TOP, Bandwidth and energy per bit for D2H and H2D data transfers, Bandwidth and energy per bit for data transfers with main memory, and Execution cycles for other functions (e.g., activation functions, normalization functions).

To model execution, the user provides a mapping scheme that assigns each layer and function to a specific hardware profile. Synchronization points can be defined where data is transferred between devices that are not interconnected. At~these points, data is sent from the last executed hardware profile to the HOST and then back to another device if needed.

\begin{figure}[!htb]
\centering
\includegraphics[width=0.35\columnwidth]{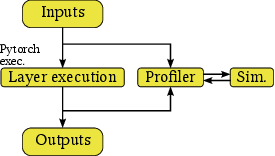}
\caption{Execution flow of the LLM hardware simulator. 
The LLM hardware simulator overrides the functions and layers of the PyTorch library.
} \label{Simulator}
\end{figure}

The simulator captures the start, end, type, and inputs of functions during graph execution (Figure~\ref{Simulator}). Based on the function type and inputs, the simulator derives all relevant metrics. Inputs are  multidimensional arrays or tensors (e.g., [batch\_size, num\_rows, num\_cols] or [batch\_size, num\_heads, num\_rows, num\_cols]).

Data transfers are modeled based on the bandwidth and energy per bit parameters of the involved hardware profiles. The total number of bits is calculated from the input data format and the metrics are derived accordingly. While data exchanges could be partially hidden by overlapping data exchanges and computations, the current baseline provides a worst-case latency scenario.

For GEMM or GEMV operations, the simulator calculates the total number of TOPs required based on the function input. It then derives the execution time and energy from the TOPS and the energy per TOP from the hardware profile. For activation or normalization functions, the simulator calculates the total number of operations required and derives the execution time and energy from the cycles and energy per operation of the hardware profile. 

Attention head computation, viewed as GEMM or GEMV operations with KV cache enabled, involves computing the current \textit{Query}, \textit{Key}, and \textit{Value} while retrieving all previous \textit{Key} and \textit{Value} results from memory. This reduces computation time at the cost of increased memory usage. The simulator accounts for these data transfers to main memory for all previous iterations in the graph along with the usual GEMM or GEMV computations.

\subsection{Hardware Profiles}
We parameterize the PIM-AI chip and DIMM to create hardware profiles for the simulator based on DDR4 PIM products such as UPMEM~\cite{UPMEMTechPaper}. In addition, we parameterize state-of-the-art NPUs from mobile SoCs: A17 Pro, Snapdragon 8 Gen 3, and Dimensity 9300, as well as the NVIDIA H100 for cloud comparisons. 

Memory energy per bit (pJ/bit) was derived from the memory technologies used: LPDDR5 for A17 Pro, LPDDR5x for Snapdragon 8 Gen 3, LPDDR5T for Dimensity 9300, and HBM for NVIDIA H100. For main memory access energy, we included both the SoC interface and the corresponding memory-side interface, resulting in a 2x factor that is not present in PIM-AI, where memory and processing units are integrated on the same chip.
Table~\ref{tab:hardware_profiles} shows the resulting hardware profiles used in this study.

For NVIDIA H100 and PIM-AI DIMMs, data transfer energy per bit include the energy of broadcasting the data to all other devices (e.g. an NVIDIA H100 within a DGX-H100 server shares the output of its sub-operation to other GPUs in the DGX-H100).
Also, we consider that GPUs within a DGX-H100 server synchronize with other GPUs through switches.
Therefore, we account 20 pJ/bit from GPU to switch and another 20pJ/bit for the opposite path.

\begin{table}[]
\caption{Hardware profiles modelled to be simulated with the LLM hardware simulator.
A PIM-AI server is composed of 24 DIMMs, each DIMM with 16 PIM-AI chips with 8 TFLOPs of computing capabilities.
A DGX-H100 server is composed of 8 H100 GPUs.
}
\label{tab:hardware_profiles}
\resizebox{\columnwidth}{!}{%
\begin{tabular}{c|cc|cc|cc|}
\cline{2-7}
                                                 & \multicolumn{2}{c|}{\textbf{Compute}}               & \multicolumn{2}{c|}{\textbf{Main memory}}                 & \multicolumn{2}{c|}{\textbf{H2D / D2H}}                   \\ \cline{2-7} 
                                                 & \multicolumn{1}{c|}{\textbf{TOPS}} & \textbf{pJ/OP} & \multicolumn{1}{c|}{\textbf{BW (GB/s)}} & \textbf{pJ/bit} & \multicolumn{1}{c|}{\textbf{BW (GB/s)}} & \textbf{pJ/bit} \\ \hline
\multicolumn{1}{|l|}{\textbf{PIM-AI chip}}       & \multicolumn{1}{c|}{5}             & 0.4            & \multicolumn{1}{c|}{102.4}              & 0.95            & \multicolumn{1}{c|}{12.8}               & 20              \\ \hline
\multicolumn{1}{|l|}{\textbf{PIM-AI server}}     & \multicolumn{1}{c|}{3072}          & 0.5            & \multicolumn{1}{c|}{39321.6}            & 0.95            & \multicolumn{1}{c|}{22 / 528}           & 1920 / 50       \\ \hline
\multicolumn{1}{|l|}{\textbf{A17 Pro}}           & \multicolumn{1}{c|}{17}            & 0.4            & \multicolumn{1}{c|}{51.2}               & 20              & \multicolumn{1}{c|}{51.2}               & 20              \\ \hline
\multicolumn{1}{|l|}{\textbf{Snapdragon 8 Gen3}} & \multicolumn{1}{c|}{17}            & 0.4            & \multicolumn{1}{c|}{77}                 & 10              & \multicolumn{1}{c|}{77}                 & 10              \\ \hline
\multicolumn{1}{|l|}{\textbf{Dimensity 9300}}    & \multicolumn{1}{c|}{16}            & 0.4            & \multicolumn{1}{c|}{76.8}               & 10              & \multicolumn{1}{c|}{76.8}               & 10              \\ \hline
\multicolumn{1}{|l|}{\textbf{DGX-H100}}          & \multicolumn{1}{c|}{7916}          & 0.5            & \multicolumn{1}{c|}{26800}              & 7               & \multicolumn{1}{c|}{450}                & 280/40          \\ \hline
\end{tabular}%
}
\end{table}

\subsection{Target Scenarios}
We envision two different deployment scenarios for LLMs:

\textbf{Cloud}: In this scenario, models can be as large as developers need them to be (e.g., larger than 7 billion parameters). The HOST has a sub-millisecond orchestration time and leverages a powerful and scalable cloud infrastructure to efficiently handle large computational loads.

\textbf{Mobile}: In this scenario, models are smaller (e.g., less or equal than 7 billion parameters~\cite{qualcomm_whitepaper}). The HOST has an orchestration period of tens of milliseconds, suitable for devices with limited computing power and energy resources.

\subsection{Benchmarks}
Open source implementations of various LLMs from Huggingface~\cite{huggingface_transformers} are used to provide a comprehensive analysis of our proposed architecture in comparison to current and widely used architectures. The simulator seamlessly integrates these pre-trained models implemented in PyTorch.

The evaluation is based on a standard experimental setup where 1000 tokens are used for the initial query (input) and 100 tokens are generated as output.

\textbf{For cloud inference}, we use the Llama2-70B and Mixtral-22x7B models, inferred using the 16-bit data format. We~normalize the number of PIM-AI DIMMs with respect to H100 GPUs in a rack unit~(U) form factor. A DGX-H100 server has an 8U form factor with 8 GPUs, while a 2U PIM-AI server has 32 DIMM slots (24 PIM-AI DIMMs and 8~standard DIMMs), allowing 96 PIM-AI DIMMs in 8U. Each model requires eight PIM-AI DIMMs, enabling parallel batch processing on 12 inference engines. Both models use 8-group GQA to manage memory requirements, and our PIM-AI architecture supports MHA because of its superior memory capacity, bandwidth, and energy efficiency.

\textbf{For mobile inference}, industry standard on-device inference techniques are applied to the Llama2-7B and Mistral-8x7B models. Both models are quantized to 4-bit weights, the KV cache is stored in 16-bit, and activation functions are computed with 16-bit.

\subsection{Performance Metrics}
We evaluate the PIM-AI architecture using specific metrics to provide a comprehensive understanding of its efficiency and performance in handling both the encoding and decoding phases of LLM operations. In addition, we include an evaluation of its overall performance to highlight its  advantages over alternative architectures.

\begin{itemize}
    \item \textbf{Encoding Performance}
    \begin{itemize}
        \item \textbf{Time to First Token}: Measures the time to produce the first token after receiving the initial prompt. Shorter times indicate better performance.
        \item \textbf{Energy Consumption}: Evaluates the total energy consumed during the encoding phase to produce the first token. Lower energy consumption indicates higher efficiency.
    \end{itemize}
    \item \textbf{Decoding Performance}
    \begin{itemize}
        \item \textbf{Tokens per second}: Measures the rate at which tokens are generated after the first token. Higher rates indicate better performance.
        \item \textbf{Energy per token}: Evaluates the energy efficiency of generating each token during the decoding phase. Lower energy per token indicates higher efficiency.
    \end{itemize}
    \item \textbf{Overall Performance}
    \begin{itemize}
        \item \textbf{Queries per Second (QPS)}: Combines time to first token and total decoding time to measure the overall rate of processing queries. Higher QPS indicates better performance.
        \item \textbf{Energy per Query (EPQ)}: Evaluates the total energy consumed per query, providing an overall measure of energy efficiency. Lower EPQ indicates better energy efficiency.
    \end{itemize}
\end{itemize}
\section{Results}
\subsection{Performance in Cloud Scenario}

\begin{figure}[!htb]
\centering
\includegraphics[width=0.95\columnwidth]{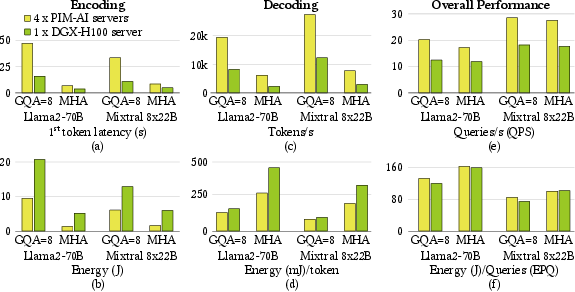}
\caption{Comparative performance of one DGX-H100 server and four PIM-AI servers. (a) Time to first token and (b) Energy consumption during the encoding phase; (c) Tokens per second and (d) Energy per token during the decoding phase; (e) Queries per Second and (f) Energy per Query as Overall Performance.} \label{fig:results_cloud}
\end{figure}

Figure~\ref{fig:results_cloud} shows the results of simulating Llama2-70B and Mixtral-8x22B models on a DGX-H100 and four PIM-AI servers. For Llama2-70B with GQA=8, batch sizes are 200 on H100 and 80 on PIM-AI; with MHA, batch sizes are 46 on H100 and 10 on PIM-AI. For Mixtral-8x22B with GQA=8, batch sizes are 200 on H100 and 80 on PIM-AI; for~MHA, batch sizes are 88 on H100 and 20 on PIM-AI. When running these models with MHA on H100 GPUs, peak performance and energy efficiency is constrained by batch size limits.

\subsubsection{Encoding Phase:}
For both the Llama2-70B and Mixtral-8x22B models, with GQA=8, the first token latency for PIM-AI is roughly 3x longer than that of the DGX-H100 server, while the energy consumption is more than halved. With MHA, the first token latency for PIM-AI is about 75\% longer than that of the DGX-H100 server, but the power consumption is about four times lower.

\subsubsection{Decoding Phase:}
For both the Llama2-70B and Mixtral-8x22B models, and for both GQA=8 and MHA configurations, PIM-AI achieves 2.23x to 2.75x more tokens per second and 15\% to 40\% less energy per token compared to the DGX-H100 server.

\subsubsection{Overall Performance:}
On average, the four PIM-AI servers process 55\% more queries per second than the DGX-H100 server, while consuming equivalent energy per query.

\subsection{Performance in Mobile Scenario}
\begin{figure}[!htb]
\centering
\includegraphics[width=0.95\columnwidth]{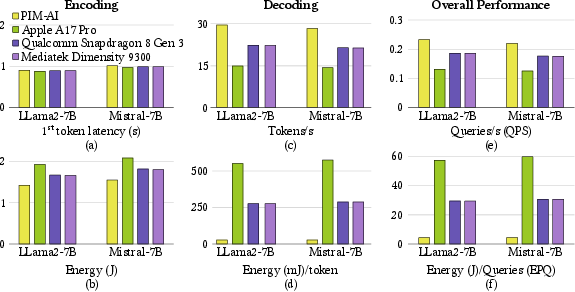}
\caption{Comparative Performance of PIM-AI on Mobile Scenario (a) Time to First Token and (b) Energy consumption during encoding phase; (c) Tokens per second and (d) Energy per token during decoding phase; (e) Queries per Second and (f) Energy per Query as Overall Performance. Each sub-figure shows PIM-AI gains over A17 Pro, Snapdragon 8 Gen3, and Dimensity 9300.} \label{fig:results_mobile}
\end{figure}

Figure~\ref{fig:results_mobile} shows the results of running the Llama2-7B and Mistral-7B models with a batch size of 1, simulating a single user interaction.

\subsubsection{Encoding Phase:}
All hardware profiles achieve similar first token latency due to their comparable TOPS capabilities. However, the PIM-AI chip achieves notable energy savings of 28.5\%, 16.4\%, and 15.3\% compared to the A17~Pro, Snapdragon 8 Gen3, and Dimensity 9300, respectively, due to its direct memory access.

\subsubsection{Decoding Phase:}
The PIM-AI chip outperforms the A17 Pro, Snapdragon 8 Gen3, and Dimensity 9300 in tokens per second due to its higher bandwidth of 102.4 GB/s compared to 77 GB/s for the other profiles. It achieves tokens per second improvements of 49.6\%, 24.5\%, and 24.7\%, respectively. The PIM-AI chip is also significantly more energy efficient, 20 times more efficient per token than the A17 Pro and 10 times more efficient than the other profiles. As the number of tokens requested increases, the superior memory bandwidth of the PIM-AI chip results in even better performance and efficiency.

\subsubsection{Overall Performance:}
PIM-AI processes around 45\% more queries per second than the A17 Pro and 25\% more than the other SoCs. It~consumes about 13.4 times less energy than the A17 Pro and 6.9 times less energy than the others.
\section{Discussion}
\subsection{Analysis of Simulation Results}
In cloud scenarios, while the initial token latency with PIM-AI is significantly higher than with the DGX-H100, the higher token generation rate compensates for this. PIM-AI achieves more queries per second for equivalent energy consumption.
The advantage of PIM-AI becomes more pronounced with longer performance. In an experimental setup with 1,000 input tokens and 1,000 output tokens, PIM-AI delivers 47\% more queries per second and consumes 15\% less energy than the DGX-H100. The primary advantage of PIM-AI in this context is a significant reduction in the 3-year total cost of ownership (TCO) per QPS, with PIM-AI showing cost ratios between 6.2x and 6.94x more favorable than the DGX-H100, depending on the LLM model. This estimate is based on a PIM-AI production server cost of approximately \$15k (\$60k for 4 servers)  compared to \$300k for a DGX-H100 server, and a world wide average electricity price of \$0.153/kWh~\cite{Electricityprices}.

For mobile devices, the overall energy efficiency of PIM-AI translates directly into a significant increase in battery life. The reduction in energy per query means that users can perform 6.9x to 13.4x more inferences before the battery runs~out. When generating 1,000 tokens per inference, these ratios are even higher at 9.8 to 19.5. 

\subsection{Limitations}
In this paper, we assume that the entire inference process, including both encoding and decoding phases, is performed on the PIM-AI. Our current PIM-AI chip significantly improves the decoding phase compared to other hardware profiles. However, the encoding phase could be optimized either by increasing the TOPS or by using a heterogeneous approach with a less energy-efficient accelerator~\cite{Samsung_HBM_PIM}. 

In addition, our LLM hardware simulator assumes constant peak performance, which simplifies modeling but does not fully capture real-world variations. This approach enables rapid prototyping and comparison of different hardware profiles. Recent research shows that mobile NPUs and high-end GPUs can achieve similar performance to our simulations~\cite{apple_models, nvidia_mistral7B}, suggesting that practical implementations may closely match our models.

\subsection{Future Perspectives}
Previous research on heterogeneous approaches has shown good performance for both encoding and decoding phases, while improving overall energy efficiency~\cite{Samsung_HBM_PIM}. Our simulations confirm that combining PIM-AI with other accelerators could optimize both phases and improve energy efficiency. However, this poses challenges for embedded devices where factors such as area, energy, power consumption, and CPU orchestration are critical. Future work will analyze this approach.

In parallel, we will validate these simulation results on a real PIM-AI chip, aiming for a prototype by the end of 2025. This validation will confirm the benefits observed in the simulations and ensure the practical applicability of the PIM-AI technology.
\section{Conclusions}
This research introduces PIM-AI, a novel PIM architecture designed to meet the computational and memory requirements of Large Language Models (LLMs). By integrating computational units into DDR5/LPDDR5 memory chips, PIM-AI addresses data transfer bottlenecks and improves performance and energy efficiency. Simulations show that PIM-AI reduces the 3-year total cost of ownership per QPS by up to 6.94x for cloud scenarios and achieves a 10x to 20x reduction in energy per token for mobile scenarios. Furthermore, PIM-AI processes 25 to 45\% more queries per second for mobile devices while consuming 6.9x to 13.4x less energy compared to state-of-the-art SoCs. These results suggest that PIM-AI can significantly enhance the efficiency, scalability, and sustainability of LLM deployments, paving the way for more advanced, energy-efficient, and cost-effective AI solutions.

\section*{Acknowledgments}
This paper is supported by the European Union’s Horizon Europe research and innovation programme (HORIZON-CL4-2021-HUMAN-01) under grant agreement No \href{https://cordis.europa.eu/project/id/101070408}{101070408}, project SustainML (Application Aware, Life-Cycle Oriented Model-Hardware Co-Design Framework for Sustainable, Energy Efficient ML Systems), and by the European Innovation Council (EIC) Accelerator and Pathfinder programmes (HORIZON-EIC-2021-ACCELERATORCHALLENGES-01 and HORIZON-EIC-2021-PATHFINDEROPEN-01-01) under grant agreement No \href{https://cordis.europa.eu/project/id/190141232}{190141232}, project ENERGIA/PIM (Accelerating Datacentre performance through Memory Chips to efficiently manage the Big Data Age) and No \href{https://cordis.europa.eu/project/id/101047160}{101047160}, project BioPIM (Processing-in-memory architectures and programming libraries for bioinformatics algorithms).

\bibliographystyle{unsrt}  
\bibliography{references}

\begin{thebibliography}{10}
\bibitem{Mistral_8x22B}
Cheaper, better, faster, stronger continuing to push the frontier of ai and making it accessible to all.
\end{thebibliography}

\end{document}